# Kinetic energy deposited into a nanodroplet, cluster, or molecule in a sticking collision with background gas

Jiahao Liang and Vitaly V. Kresin[*]

*Department of Physics and Astronomy, University of Southern California,*
*Los Angeles, California 90089-0484, USA*

**Abstract**

In processes when particles such as nanodroplets, clusters, or molecules move through a dilute background gas and undergo capture collisions, it is often important to know how much translational kinetic energy is deposited into the particles by these pick-up events. For sticking collisions with a Maxwell-Boltzmann gas, an exact expression is derived which is valid for arbitrary relative magnitudes of the particle and thermal gas speeds.

---

[*] Corresponding author. Email: *kresin@usc.edu*

*Introduction.* A well-known way to add atoms or molecules to a free nanocluster or nanodroplet in a beam is the pick-up method (see, e.g., the reviews[1,2]). The beam passes through a cell filled with a dilute vapor of the substance of interest, and one or more atom or molecule collides with and sticks to the cluster.

This is a perfectly inelastic collision, and oftentimes one needs to know how much kinetic energy it deposits into the cluster. For example, the great majority of helium nanodroplet experiments involve pick-up doping, and the accompanying energy release is an omnipresent consideration. Nanodroplets dissipate this energy by prompt evaporation of He atoms from the surface, and their resulting shrinkage affects both the dopant's environment and the probability of a subsequent pick-up event.[3-5] For metal and water clusters, analogous evaporative attachment processes have been used to assess their caloric curves.[6,7]

The dilute vapor in the pick-up cell can be treated as an ideal gas. Its temperature $T$ can vary from approximately room temperature for volatile molecular dopants up to ~2000 K for high melting point metals.[4,8] This corresponds to molecular thermal speeds $v$ ranging roughly from $10^2$ to $10^3$ m/s. Since the beam speed $\mathcal{V}$ can vary over a similar range, depending on the cluster size and the source conditions, we can encounter situations with very different ratios between $v$ and $\mathcal{V}$.

A commonly encountered estimate of the deposited translational energy is $\langle E_t \rangle \approx \tfrac{3}{2} k_B T + \tfrac{1}{2} m \mathcal{V}^2$ where $m$ is the molecular mass.[6,9] (Sometimes the reduced mass correction is included.) This is inferred from the familiar expression for the mean energy in the Maxwell-Boltzmann distribution, but is actually incorrect for a binary collision. For example, in the case of a particle at rest ($\mathcal{V}=0$) immersed in a gas at temperature $T$, the average translational kinetic energy deposited per impact is $2k_B T$ (see below). The reason for the larger prefactor is that the flux of molecules is proportional to their velocity, and therefore the target experiences more hits from the higher-energy part of the speed distribution.

When the particle is moving, the situation grows more complex because there is now an additional shift in the population of molecules able to collide with it. For some molecules the relative speed is now higher, for some it is lower, and some are too slow to catch up to the particle at all. This varies not only with the magnitude of the molecular velocity but also with its direction. In addition, since the ratio between $v$ and $\mathcal{V}$ may cover a wide range, one cannot simply transform the Maxwell-Boltzmann velocity distribution into the beam reference frame and expand it in powers of the speed ratio.[10] Therefore a careful evaluation of the distribution of relative velocities and the corresponding fluxes is necessary. In this note we provide an exact expression for the translational energy deposition in this general case. It is based on extending a calculation described in treatises on the kinetic theory of gases. As such, it is possible that the result has been discussed elsewhere in a different context. However, to the best of our knowledge it has not been presented or applied in the analysis of molecular beam experiments with nanodroplets and nanoclusters, and we believe that it may be of use for such work.

*Stationary particle.* Let us begin by reviewing the aforementioned case of a stationary particle surrounded by a gas of molecules at a temperature $T$. The average translational energy deposited by a captured molecule can be evaluated as the ratio of the kinetic energy flow into the particle to the average number of collisions. If $S$ is the particle's surface, then the number of impacts it experiences per unit time is given by the total flux of gas molecules through $S$:

$$N_m = n \int d^3v\, P(\vec{v}) \oiint_S \vec{v} \cdot d\vec{a} = \left[ n \int dv\, f(v) v^3 \right] \cdot \left( \oiint_S \hat{v} \cdot d\vec{a} \right). \tag{1}$$

Here $n$ is the number density of molecules in the gas, $P(\vec{v})$ is the probability distribution of their velocities, and $\hat{v}$ is a unit vectors such that $\vec{v}=v\hat{v}$. Since the directions of $\hat{v}$ are distributed randomly, the expression separates into an integral over the Boltzmann distribution $f(v)\propto\exp(-mv^2/2k_BT)$ and a purely geometrical surface integral factor.

Analogously, the translational kinetic energy influx is

$$\Phi_t = \left[n\int dv\left(\tfrac{1}{2}\mu v^2\right)f(v)v^3\right]\cdot\left(\oiint_S \hat{v}\cdot d\vec{a}\right), \quad (2)$$

where $\mu$ is the reduced mass of the particle-molecule system. For simplicity, we assume an energy-independent sticking coefficient of unity. With the Boltzmann $f(v)$, the average energy deposited in one collision is easily calculated to be $\langle E_t\rangle = \Phi_t/N_m = 2k_BT(\mu/m)$. If $m\ll M$, where $M$ is the particle mass, then $\mu/m\approx 1$ and we recover the simple result mentioned earlier. It is of course identical to the well-known fact[11,12] that the classical average thermal kinetic energy of molecules, electrons, etc. evaporating from, or passing through, a surface is $2k_BT$.

*Moving particle.* The purpose of the preceding outline is to suggest that in order to compute the quantity $E_t$ for a moving particle we can employ the same procedure: find the ratio of the energy flux delivered by the gas molecules into the particle to the collision rate. As stated above, this task is more complex when the particle is moving with respect to the Maxwell-Boltzmann gas.

It turns out that the problem is very closely related to one encountered in the calculation of the mean free path of a molecule of a given speed in the presence of all the other gas molecules. The latter was found a long time ago and is referred to as the Tait mean free path.[13] It is described in detail in classic books on kinetic theory,[14-16] where the application to momentum transfer in elastic intermolecular collisions is also discussed. Here, we make use of this approach to evaluate the energy transfer in *in*elastic collisions.
4

The idea is to consider how many collisions per unit time a particle moving at $\vec{\mathcal{V}}$ will undergo with molecules having a particular velocity $\vec{v}$. It is given by the number of the latter molecules contained within a cylinder with base equal to the cluster-molecule collision cross section $\sigma$ and height equal to their relative velocity $V_r$. If the angles between $\vec{\mathcal{V}}$ and $\vec{v}$ are $\theta$ and $\varphi$, then the corresponding number of collisions per unit time is

$$dN_m = n\sigma V_r f(v) v^2 dv \sin\theta d\theta d\varphi. \tag{3}$$

Since $V_r^2 = \mathcal{V}^2 + v^2 - 2\mathcal{V}v\cos\theta$, Eq. (3) can be rewritten as follows, after integrating over $\varphi$:

$$dN_m = 2\pi n\sigma \frac{v}{\mathcal{V}} V_r^2 f(v) v^2 dv dV_r. \tag{4}$$

The total number of collisions per unit time is obtained by first integrating over the allowed values of $V_r$ (as defined by the range $0 \le \theta \le \pi$) and then over $v$ from zero to infinity. The integration limits of $V_r$ require care, as they depend on whether $v < \mathcal{V}$ or $v > \mathcal{V}$, and therefore the final integration over $v$ will be performed separately over the ranges $0 \le v \le \mathcal{V}$ and $\mathcal{V} \le v \le \infty$. The details are written out in the books cited above, and the final result is

$$N_m = \frac{1}{\sqrt{\pi}} v_p n\sigma \frac{\Psi(x)}{x}, \tag{5}$$

where $x \equiv \mathcal{V}/v_p$ and $v_p$ is the most probable speed of the molecular gas:

$$v_p = \sqrt{\frac{2k_B T}{m}}. \tag{6}$$

The function $\Psi$ is given by

$$\Psi(x) = xe^{-x^2} + \sqrt{\pi}\left(\tfrac{1}{2} + x^2\right)\mathrm{erf}(x) \tag{7}$$

with $\mathrm{erf}(x) = \left(2/\pi^{1/2}\right)\int_0^x \exp(-y^2)dy$.

Now we need to determine the amount of kinetic energy delivered by the colliding molecules per unit time. The calculation proceeds in the same way as above, with the addition of a kinetic energy factor to Eq. (4):

$$d\Phi_t = \left(\tfrac{1}{2}\mu V_r^2\right) 2\pi n\sigma \frac{v}{\mathcal{V}} V_r^2 f(v) v^2 dv dV_r \tag{8}$$

We can perform the integration over $V_r$ and $v$ in the same way as described above. Then, dividing the energy flux by the collision rate we finally obtain an expression for the average amount of kinetic energy deposited into the particle by one molecular collision:

$$\langle E_t \rangle = k_B T (\mu/m) \frac{\Theta(x)}{\Psi(x)}, \tag{9}$$

where the function $\Psi(x)$ is defined in Eq. (7) and

$$\Theta(x) = x\left(\tfrac{5}{2} + x^2\right)e^{-x^2} + \sqrt{\pi}\left(\tfrac{3}{4} + 3x^2 + x^4\right)\mathrm{erf}(x). \tag{10}$$

Eq. (9) is the main result of this paper.

*Limiting cases.* Let us consider the limiting cases. If the nanodroplet/cluster beam is moving much slower than the molecules it picks up ($x \ll 1$), an expansion of Eq. (9) gives

$$\langle E_t \rangle \approx 2k_B T(\mu/m)\left(1 + \tfrac{2}{3}x^2\right) \tag{11}$$

which reduces to the limit discussed above for $x=0$. On the other hand, if the beam is much faster than the molecules' thermal motion, then an expansion of Eq. (9) for $x \gg 1$ brings it to the form

$$\langle E_t \rangle \approx \frac{\mu}{m}\left(\tfrac{1}{2}m\mathcal{V}^2 + \tfrac{5}{2}k_B T\right) \tag{12}$$

This agrees with the limit stated in ref.[7]

*Simulation.* As an illustration, we carried out a molecular dynamics simulation of collisions experienced by a particle in uniform motion through a thermal gas. A volume of identical gas molecules interacting by pairwise Lennard-Jones potentials was randomly initialized in LAMMPS[17,18] according to the Maxwell-Boltzmann distribution. While the simulation was running, the position and velocity data were saved to a file every five timeframes. In the second step, a MATLAB script was created to simulate a massive spherical particle moving at constant velocity through this volume. If a gas molecule entered a predefined collision area around the particle, this was recorded as a capture event and the kinetic energy lost by the molecule was computed. The simulation was repeated a large number of times for different particle velocities. The result is shown in Fig. 1 and is in perfect agreement with the analytical expression.

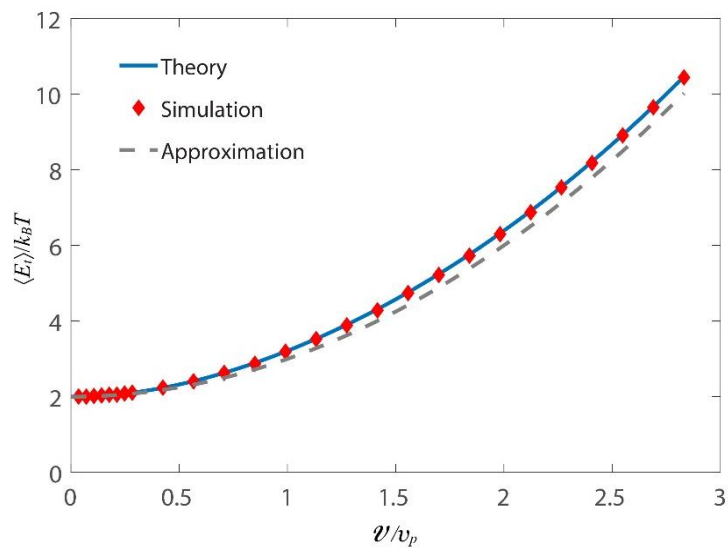

**Figure 1**. Average translational energy $\langle E_t \rangle$ deposited in a pick-up collision between a particle moving at a uniform velocity $\mathcal{V}$ and a molecule from the surrounding gas with temperature $T$ and the most probable speed $v_p$. Diamonds: molecular-dynamics simulation, solid line: Eq. (9). The dashed line is the approximation $\langle E_t \rangle \approx 2k_B T + \frac{1}{2} m \mathcal{V}^2$ (For this plot the particle is assumed to be heavy, so that $\mu = m$.)

*Concluding remarks.* The derivation described here assumes a velocity-independent capture cross section. If the cross section is instead velocity-dependent, for example in cases where strong long-range polarization or dispersion forces are important, the result can be generalized in a straightforward manner by incorporating an additional power of the relative velocity $V_r$ in Eqs. (3) and (8). This will be considered elsewhere.

Furthermore, the constant-velocity case considered here can be extended to a variable velocity $\vec{\mathcal{V}}(t)$. This would be applicable to situations when clusters/nanodroplets are gradually slowed down by successive collisions, or to particles in ion or optical traps and molecules in Stark or Zeeman slowers.

*Acknowledgment.* This work was supported by the U.S. National Science Foundation under Grant No. CHE-1664601.

*Data availability.* Data sharing is not applicable to this article as no new data were created or analyzed in this study.